\documentclass[traditabstract,printer]{aa}
\usepackage{graphicx}
\usepackage{epsfig}
\usepackage{amssymb,amsmath}
\usepackage{natbib}
\bibpunct{(}{)}{;}{a}{}{,}
\usepackage{pslatex}
\usepackage{caption}
\usepackage{longtable}
\usepackage{hyperref}
\usepackage{gensymb}

\begin{document}

\title{Gravitational waves from post-merger radially-oscillating millisecond pulsars}

\author{Z. G. Dai$^{1,2}$\thanks{E-mail: \href{dzg@nju.edu.cn}{dzg@nju.edu.cn}}}

\institute{$^1$School of Astronomy and Space Science, Nanjing University, Nanjing 210093, China; \\
$^2$Key Laboratory of Modern Astronomy and Astrophysics (Nanjing University), Ministry of Education, China}

\authorrunning{Dai}
\titlerunning{Gravitational waves from newborn millisecond pulsars}

\abstract{
Observations of short-duration gamma-ray bursts and their afterglows show that a good fraction (perhaps $\gtrsim50\%$) of binary neutron star mergers lead to strongly magnetized, rapidly rotating pulsars (including millisecond magnetars), no matter whether the pulsar remnants are short- or long-lived. Such compact objects are very likely to have significant radial oscillations and high interior temperatures, as indicated in recent numerical simulations. In this paper, we have investigated rotation-induced gravitational radiation from possibly existing, radially oscillating pulsars after binary neutron star mergers, and find that this mechanism can efficiently damp the radial oscillations. The resulting gravitational waves (GWs) could have a non-negligible contribution to the high-frequency spectrum. We provide an order-of-magnitude estimate of the event rate and suggest that such GW events would be detectable with the advanced LIGO/Virgo or next-generation detectors. Our discussion can also be applied to newborn, radially oscillating, millisecond pulsars formed through the other astrophysical processes.}


\maketitle

\section{Introduction}

Plenty of indirect evidence shows that binary neutron star mergers are the origin of short gamma-ray bursts (GRBs) \citep[for reviews see][]{nakar07,berger14}. Direct evidence comes from the recently-discovered gravitational wave (GW) event GW170817 and its electromagnetic counterparts \citep{Abbott17a,Abbott17b}. Post-merger central objects include two possibilities: black holes and strongly-magnetized millisecond pulsars, which are dependent on the total mass of a binary neutron star system and the equation of state for neutron-star matter.

The central millisecond pulsar engine (including millisecond magnetars) was proposed as the predicted temporal plateaus of GRB afterglows by taking rotational energy injection into account for post-burst relativistic blast waves \citep{Dai98a,Dai98b,zhang01,Dai04,Dall11}, and to explain late-time X-ray flares in short GRBs by considering magnetic reconnection events taking place over the surfaces of post-merger millisecond pulsars \citep{Dai06}. This kind of engine is not only consistent with general relativistic hydrodynamic simulations of binary neutron star mergers \citep{Giacomazzo13,Piro17} but also with a large sample of short GRB afterglows with temporal plateaus \citep{Rowlinson10,Rowlinson13}. A statistical analysis shows that such an engine occurs in a good fraction (perhaps $\gtrsim50\%$) of short GRB afterglows \citep{gao16,Piro17}. For GW170817, a neutron star remnant is suggested to interpret an observed multiwavelength kilonova \citep{yu18,ai18,li18,metzger18}, a broadband longevous afterglow \citep{geng18}, and a possible X-ray flare appearing at $\sim 155\,$days after this merger event \citep{piro18}.

Recent numerical simulations indicate radial oscillations in post-merger neutron star remnants \citep[e.g.,][]{bau18}. The remnants have a radial-pulsation amplitude $\alpha\sim 0.03-0.1$ and a high temperature $T\gtrsim 10-20\,{\rm MeV}/k$ (where $k$ is the Boltzmann constant). For these rapidly rotating neutron stars, dynamical bar-mode and secular instabilities are unlikely to take place because the ratio of the stellar rotational energy to gravitational binding energy is below $0.14$ for many realistic neutron-matter equations of state \citep{Piro17}. The remaining spin-related instabilities such as unstable r-modes \citep{andersson98,owen98,ho00,Bond09,Bond13,Dai16} and higher multipole (including quadrupolar) f-modes \citep{pas13} are attributed to $l=m=2$ and $l=m>2$ non-radial pulsations respectively. However, their pulsation amplitudes necessarily spend a very long time -- at least a few hundreds of seconds -- increasing from extremely small to saturation values. Therefore, these instabilities cannot affect the rotational evolution of newborn millisecond pulsars with age $\lesssim 1\,$s.

In this paper, we investigate GWs from post-merger rapidly-rotating radially-oscillating neutron stars and show that such GW events would be detectable with the advanced LIGO/Virgo or next-generation detectors. Although GWs from binary neutron star mergers have been presented through numerical simulations \citep[e.g.,][]{ber15,bau15,tak15,chat17,shibata17,zappa18}, GWs from post-merger radially-oscillating neutron stars remain unexplored\footnote{After this paper was submitted, the prospects of studying GW170817-like post-merger signals with future GW detectors were investigated by \cite{torres18}.}. Interestingly, \cite{Wein13} and \cite{andersson18} explored dynamical tidal effects on GWs from inspiraling neutron star binaries before the mergers, and \cite{Dall15} studied GWs from post-merger neutron stars by considering mass quadrupole moments that are induced by magnetic field amplification in the interiors during the mergers. In order to assess the importance of GWs from post-merger radially-oscillating neutron stars, we propose a physical model and find that they could have a non-negligible contribution to the post-merger spectrum at a frequency $\gtrsim3\,$kHz. We also suggest that this model can be used to discuss GWs from newborn, radially oscillating, millisecond pulsars formed through other astrophysical processes.

This paper is organized as follows. We first analyze three damping mechanisms of the radial oscillations, described in Section 2, and then we discuss the detectability of the resulting GWs in Section 3. We apply this analysis to GW170817 in Section 4, and discuss some implications of our model and give an order-of-magnitude estimate of the event rate in Section 5. Finally, we summarize our conclusions in Section 6.

\section{Damping mechanisms}
We consider a model in which a radially oscillating, strongly magnetized, rapidly rotating pulsar occurs just after the merger of two neutron stars. At this moment, any non-radial hydrodynamical effect is insignificant so that the pulsar remnant has a radial oscillation alone, as shown in recent numerical simulations \citep{bau18}. We also neglect the effect of any non-radial instability (e.g., unstable r-mode and f-mode) because its initial pulsation amplitude is extremely small so that the stellar rotation remains unchanged for $t\lesssim1\,$s \citep{andersson98,owen98,ho00,Bond09,Bond13,pas13,Dai16}. The radial oscillation energy of this neutron star is calculated by \citep{chau67,sawyer89}
\begin{eqnarray}
{\cal E} & = & \frac{3}{20}MR^2\alpha^2\omega^2=
0.75\times 10^{51}\left(\frac{M}{2.5M_\odot}\right)R_6^2
\alpha_{-1}^2\omega_4^2\,{\rm erg},
\end{eqnarray}
where $M$ and $R$ are the stellar mass and radius respectively, $\alpha\equiv \Delta R/R$ is the radial pulsation amplitude,
$\omega$ is the angular pulsation frequency, and the convention $Q_x=Q/10^x$ is adopted in cgs units. By linearizing the differential hydrodynamic equations describing the stellar non-equilibrium mechanical behavior and considering only small-amplitude adiabatic oscillations, the angular pulsation frequency turns out to be \citep{cox80}
\begin{eqnarray}
\omega & = & \left[\frac{4\pi}{3}(3\gamma-4)G\rho\right]^{1/2}\\
& = & 2.6\times 10^4\eta_\gamma^{1/2}\left(\frac{M}{2.5M_\odot}\right)^{1/2}R_6^{-3/2}\,{\rm s}^{-1},
\end{eqnarray}
where $\gamma$ is the adiabatic index of neutron-star matter, $\eta_\gamma\equiv (3\gamma-4)/2$, and $\rho=M/(4\pi R^3/3)$ is the stellar average mass density. Inserting Eq. (3) into Eq. (1), we obtain
\begin{eqnarray}
{\cal E} = 5.1\times 10^{51}\eta_\gamma\left(\frac{M}{2.5M_\odot}\right)^{2}R_6^{-1}\alpha_{-1}^2\,{\rm erg}.
\end{eqnarray}

This radial oscillation energy is lost through three mechanisms, which we discuss in detail below. The first damping mechanism is rotation-induced gravitational radiation \citep{chau67}, which has been used to investigate GWs from phase transitions of accreting neutron stars in low-mass X-ray binary systems \citep{cheng98}. The GW luminosity via this mechanism is given by
\begin{eqnarray}
\dot{\cal E}_{\rm GW} & = & \frac{1}{375}\frac{G}{c^5}
\left(\frac{45}{4}\gamma-\frac{9}{5}\right)^2\left(\frac{\Omega}{\omega}\right)^4M^2R^4\alpha^2\omega^6 \\
& = & 0.81\times 10^{54}\eta_\gamma\kappa_\gamma^2\left(\frac{M}{2.5M_\odot}\right)^3R_6\alpha_{-1}^2P_{-3}^{-4}\,{\rm erg}\,{\rm s}^{-1},
\end{eqnarray}
where $\Omega=2\pi/P$ is the stellar angular rotation frequency with $P$ being the rotational period \citep{chau67}, and $\kappa_\gamma\equiv (225\gamma-36)/414$. Therefore, the corresponding damping timescale is
\begin{eqnarray}
t_{\rm GW}\equiv \frac{{\cal E}}{\dot{\cal E}_{\rm GW}}
=6.3\kappa_\gamma^{-2}\left(\frac{M}{2.5M_\odot}\right)^{-1}R_6^{-2}P_{-3}^{4}\,{\rm ms}.
\end{eqnarray}

The second damping mechanism is pulsational magnetic radiation (PMR) due to a temporally-changing magnetic dipole moment $|{\bf m}|=BR^3/2\propto R(t)$, where $B$ is the stellar surface field strength at the magnetic pole \citep{hoyle64,cameron65,heintz72,duncan89}. The resulting electromagnetic emission power is written as\footnote{We note that the factor multiplying $c^3$ in the denominator of the right term of the second equality sign should be six rather than twelve in \cite{duncan89}.}
\begin{eqnarray}
\dot{\cal E}_{\rm PMR} & = & \frac{2|{{\bf \ddot{m}}}|^2}{3c^3} = \frac{B^2R^6\alpha^2\omega^4}{6c^3}\nonumber \\
& = & 2.8\times 10^{47}\left(\frac{M}{2.5M_\odot}\right)^2B_{14}^2\alpha_{-1}^2\,{\rm erg}\,{\rm s}^{-1},
\end{eqnarray}
and the corresponding damping timescale is
\begin{eqnarray}
t_{\rm PMR}\equiv \frac{{\cal E}}{\dot{\cal E}_{\rm PMR}}=1.8\times 10^4\eta_\gamma B_{14}^{-2}R_6^{-1}\,{\rm s}.
\end{eqnarray}
This timescale is much larger than $t_{\rm GW}$ for any reasonable surface field strength (i.e., $B\lesssim10^{16}\,$G).

The third damping mechanism is the bulk viscosity of neutron-star matter \citep{sawyer80,sawyer89}. At the initial stage post merger, the newborn pulsar has a high temperature of $T\gtrsim 10-20\,{\rm MeV}/k$, at which neutrinos are trapped in the interior and form an ideal Fermi-Dirac gas with chemical potential $\mu_\nu\gg kT$ \citep{sawyer79}. As a result, the radial pulsations are damped through the following non-equilibrium reactions: $p+e^-\leftrightarrow n+\nu_e$. According to these reactions, \cite{Reisen92} estimated the viscous damping timescale for freely escaping neutrinos (see their Eq. [37]), which is certainly much larger than $t_{\rm GW}$. \cite{cheng98} derived the bulk viscosity coefficient ($\zeta$) for trapped neutrinos and the corresponding damping timescale\footnote{The chiral angle $\theta$ in Eq. (5) of \cite{cheng98} is taken to be zero in this paper, which implies no kaon condensation.}
\begin{eqnarray}
t_{\rm V,N} \equiv \frac{\rho R^2}{30\zeta}=22Y_e^{1/3}Y_\nu^{-2/3}Y_n^{-4/3}\rho_{15}^{-2/3}R_6^2\left(\frac{kT}{10{\rm MeV}}\right)^2\,{\rm s},
\end{eqnarray}
where $Y_e$, $Y_\nu$, and $Y_n$ are respectively the fractions of electrons, neutrinos, and neutrons. In the following, we calculate these fractions and $\Lambda(l_0,n_b)\equiv Y_e^{1/3}Y_\nu^{-2/3}Y_n^{-4/3}$ to obtain $t_{\rm V,N}$. For a neutral mixture of free non-relativistic neutron, non-relativistic proton, relativistic electron and relativistic neutrino gases in the interior of the post-merger neutron star, if the lepton fraction is taken to be $l_0$, then we have $Y_\nu=l_0-Y_e$ as well as the proton fraction $Y_p=Y_e=1-Y_n$. The chemical equilibrium condition leads to $\xi Y_e^{1/3}+Y_p^{2/3}=Y_n^{2/3}+\xi (2Y_\nu)^{1/3}$, where the factor of two multiplying $Y_\nu$ in the last term is due to the existence of only one coupled polarization state for $\nu_e$ and $\xi\equiv (3\pi^2)^{-1/3}n_b^{-1/3}(2m_nc)/\hbar=3.1(n_b/n_0)^{-1/3}$ with $n_b$ being the baryon number density and $n_0$ being the nuclear baryon number density \citep{sawyer80}. Therefore, we can calculate the fractions of four kinds of particles and then $\Lambda(l_0,n_b)$ for given $l_0$ and $n_b$. Figure 1 presents $\Lambda(l_0,n_b)$ as a function of $n_b$ for $l_0=0.07$, $0.1$, and $0.15$. We can see from this figure that $\Lambda(l_0,n_b)>10$, and thus from Eq. (10) that the viscous damping timescale $t_{\rm V,N}$ is also much larger than $t_{\rm GW}$.

\begin{figure}
\begin{center}
\includegraphics[width=0.53\textwidth,angle=0]{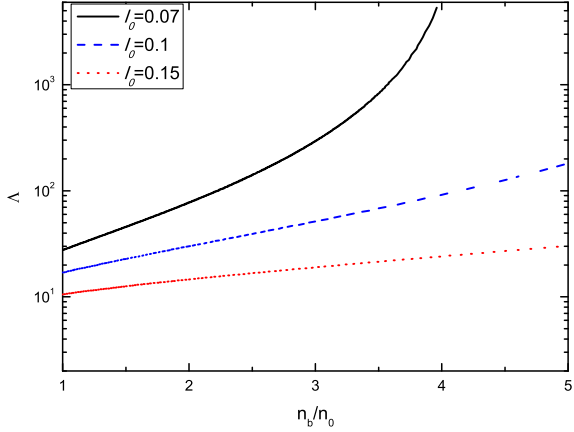}
\caption{$\Lambda(l_0,n_b)\equiv Y_e^{1/3}Y_\nu^{-2/3}Y_n^{-4/3}$ as a function of $n_b/n_0$ in Eq. (10) for different values of $l_0$. The black, blue, and red lines are corresponding to $l_0=0.07$, $0.1$, and $0.15$, respectively.} \label{fig1}
\end{center}
\end{figure}

An alternative possibility for the central object post merger is that it is a strange quark star. This possibility was put forward by \cite{Dai98b}, recently implied from a statistic analysis of the observed plateau durations in the light curves of short GRB afterglows by \cite{li16}, and very recently suggested in numerical simulations by \cite{most18} and \cite{bau18}. In this case, the bulk viscosity arises from non-equilibrium non-leptonic reactions among quarks: $u+d\leftrightarrow u+s$, through which the viscous damping timescale of radial oscillations in the strange quark star is estimated by
\begin{eqnarray}
t_{\rm V,Q}=12\rho_{15}^2R_6^2\left(\frac{m_sc^2}{100{\rm MeV}}\right)^{-4}\left(\frac{kT}{10{\rm MeV}}\right)^2\,{\rm s},
\end{eqnarray}
where $m_s$ is the strange quark mass \citep{dai96,madsen92,madsen00}. This timescale is still far beyond $t_{\rm GW}$ for typical values of the relevant parameters (e.g., $\rho_{15}\sim 1$, $R_6\sim 1$, $m_sc^2\sim 100\,$MeV, and $kT\gtrsim 10\,$MeV).

Therefore, we can conclude that no matter whether the post-merger radially oscillating object is a neutron star or a strange quark star, the damping mechanisms due to pulsational magnetic radiation and bulk viscosity are both negligible and thus the gravitational radiation must be the dominant damping mechanism. In the next section, we discuss the detectability of such GWs.

\section{Detectability of GWs}

The strain of GWs can be estimated through using the mass quadrupole approximation to the Einstein field equations \citep{Shapiro83,thorne87}. This approximation shows that the GW strain is given by
\begin{eqnarray}
h\simeq \frac{2G}{c^4}\frac{|\ddot{Q}|}{d},
\end{eqnarray}
where $d$ is the distance to the source \citep{Shapiro83}, and $Q$ is the time-dependent mass quadrupole moment \citep{chau67}
\begin{eqnarray}
Q=\frac{8\pi}{15}\left(\frac{45}{4}\gamma-\frac{9}{5}\right)\left(\frac{\Omega}{\omega}\right)^2\rho\alpha R^5\cos(\omega t).
\end{eqnarray}
Inserting Eq. (13) into Eq. (12), we have
\begin{eqnarray}
h\simeq 0.88\times 10^{-22}\kappa_\gamma\alpha_{-1}P_{-3}^{-2}R_6^2\left(\frac{M}{2.5M_\odot}\right)\left(\frac{d}{100{\rm Mpc}}\right)^{-1}.
\end{eqnarray}

The characteristic GW strain can be approximated by \citep{Corsi09}
\begin{eqnarray}
h_c=fh\sqrt{\frac{dt}{df}}\simeq h\sqrt{ft_{\rm GW}},
\end{eqnarray}
where $f=\omega/2\pi$ is the frequency of GWs. Inserting Eqs. (3), (7), and (14) into Eq. (15), we find
\begin{eqnarray}
h_c\simeq 4.5\times 10^{-22}\eta_\gamma^{1/4}\alpha_{-1}R_6^{1/4}\left(\frac{M}{2.5M_\odot}\right)^{3/4}\left(\frac{d}{100{\rm Mpc}}\right)^{-1}.
\end{eqnarray}
This equation shows that $h_c$ is not only independent of the neutron-star rotational period but also weakly dependent of the stellar radius. In addition, the inferred remnant mass $M$ is generally close to $2.5M_\odot$ for most of the observed binary neutron star systems in the Galaxy \citep[also see Table 1 of][]{baiotti17}. Thus, the radial pulsation amplitude $\alpha$ could be found if $h_c$ is detected and if $d$ is inferred at the inspiraling stage. On the other hand, if no GW signal from the radial oscillations is detected, we can give an upper limit on $\alpha$ by comparing $h_c$ with the sensitivity (i.e., noise level) of the GW detector, $h_{\rm rms}=[fS_n(f)]^{1/2}$, where $S_n(f)$ is the power spectral density (PSD) of the detector noise. The PSD has been presented as a function of GW frequency for the advanced LIGO detector and the Einstein Telescope (ET) respectively \citep[e.g.,][]{arun05,mishra10,sun15,gao17}.

\section{Application to GW170817}

The observations of GW170817 indicate that this event should have arisen from an inspiral of two neutron stars, the total mass of which is $M_{\rm tot}\simeq 2.74M_\odot$ \citep{Abbott17a}. If the two stars are assumed to have the same mass, then the gravitational mass of each star becomes $\simeq1.37M_\odot$. Because the baryonic mass ($M_b$) and gravitational mass ($M_g$) of a neutron star satisfy a correlation, $M_b=M_g+0.075M_g^2$ \citep{timmes96}, the total baryonic mass of the binary system turns out to be $M_{b, \rm tot}\simeq 3.02M_\odot$. In addition, the baryonic mass of the ejecta during the merger is found to be $M_{\rm ej}\simeq 0.065M_\odot$ by fitting the multiwavelength kilonova data \citep{villar17}, so the gravitational mass of the neutron star left behind after the merger is derived as $M_{g, \rm tot}\simeq 2.49M_\odot$. This is why the neutron star remnant mass $M$ in the two sections above is scaled as $2.5M_\odot$. The actual gravitational mass of the neutron star remnant is dependent of the mass ratio ($q$) of the pre-merger two neutron stars. However, since $q\sim 1$ for most of the observed binary neutron star systems in the Galaxy \citep{baiotti17}, the mass $M$ of each neutron star remnant from these systems is inferred to be around $2.5M_\odot$. Furthermore, since $h_c$ in Eq. (16) is weakly dependent of the stellar radius, the remnant structure would scarcely influence our theoretical GW strain if $\alpha$ is fixed.

The advanced LIGO/Virgo detectors searched for GWs from the neutron star remnant after GW170817 but no signal was found \citep{Abbott17c,Abbott18}. Recently, \cite{vanPutten18} claimed the detection of a post-merger signal candidate with a duration  $\sim 1\,$s but their estimated GW energy is lower than the sensitivity estimates of \cite{Abbott17c}. The upper limits on the GW strain were derived for two different observed periods of a GW signal by \cite{Abbott17c}. For example, the best upper limit on the root-sum-square of the GW strain emitted from $1-4$\,kHz, for a signal $\lesssim 1\,$s, is $h^{50\%}_{\rm rss}=2.1\times 10^{-22}\,{\rm Hz}^{-1/2}$ at the $50\%$ detection efficiency \citep{Abbott17c}. From Eq. (16), we find that the strain of GWs from the neutron star remnant after GW170817 is $h_c\simeq 1.1\times 10^{-21}\alpha_{-1}$ for $M\simeq 2.5M_\odot$, $R_6\simeq 1$ and $d\simeq 40\,$Mpc (hereafter $\gamma=2$ is assumed so that $\eta_\gamma=\kappa_\gamma=1$). The requirement that $h_c\lesssim \sqrt{f}h^{50\%}_{\rm rss}$ leads to a radial pulsation amplitude $\alpha\lesssim0.6\sqrt{f_3}$. Unfortunately, this limit seems too loose to provide useful information. Post-merger GW emission from a similar event would be possibly detectable with next-generation detectors such as ET or when the advanced LIGO/Virgo detectors reach their design sensitivity \citep{Abbott17c}. Once detected, a sample of such GW events would provide a unique probe for post-merger central objects.

\section{Discussion}

Gravitational waves from post-merger radial oscillations may have a non-negligible contribution to the high-frequency spectrum. On the one hand, such GWs are emitted at a frequency $f\simeq 4.1R_6^{-3/2}(M/2.5M_\odot)^{1/2}\,$kHz from Eq. (3), which is very close to the cutoff frequency shown in Figure 1 of \cite{chat17}. On the other hand, the characteristic strain $h_c\simeq 4.5\times 10^{-22}\alpha_{-1}R_6^{1/4}(M/2.5M_\odot)^{3/4}$ from Eq. (16) is much stronger than the numerical peak strain $h_p\sim 10^{-24}$ at a frequency $\gtrsim3\,$kHz in \cite{chat17} if the distance to the source is 100\,Mpc. In fact, GWs from post-merger spin-related non-radial instabilities must be radiated at a frequency that is about twice as large as the Kepler rotation limit of the neutron star remnants, that is, $f\simeq 2/P_{\rm Kepler}\simeq 3.5R_6^{-3/2}(M/2.5M_\odot)^{1/2}\,$kHz, where $P_{\rm Kepler}$ is the stellar Kepler rotational period \citep{haensel09}. Thus, the frequency of GWs from radial oscillations is nearly equal to that from non-radial instabilities. Moreover, the pulsation amplitudes of non-radial instabilities must grow to some saturation values in a long period of at least a few hundreds of seconds \citep{andersson98,ho00,Dai16,pas13}, so that GWs from them are much weaker than those from the radial oscillations for $t\lesssim1\,$s, as shown by comparing $h_c$ and $h_p$.

Observationally, a GW detector would first discover GWs from an inspiral of two neutron stars and derive the masses of the two stars and the distance to the source, as in GW170817. If the detector subsequently discovers GWs from the post-merger central object and if such GWs are due to radial oscillations of a millisecond pulsar remnant, then we can constrain the radial pulsation amplitude $\alpha$ from Eq. (16), provided that $h_c$ is detected and that $d$ is inferred from the GWs at the inspiraling stage. In addition, the frequency and duration of post-merger GWs would be found from their waveforms. Furthermore, according to Eqs. (3) and (7), one would obtain the mass-radius relation and rotational period of the pulsar remnant, which, together with the inferred pulsation amplitude $\alpha$, would provide useful information for constraining the post-merger central object.

We give here an order-of-magnitude estimate of the detectable GW event rate. From \cite{Abbott17a}, the best rate of GW170817-like events is approximated by ${\cal R}_{\rm tot}\sim1540\,{\rm Gpc}^{-3}\,{\rm yr}^{-1}$. To be conservative, we assume that a half of such events can produce neutron star remnants \citep{gao16,Piro17}, so the rate of GW events from radial oscillations is ${\cal R}_{\rm osc}\sim 800\,{\rm Gpc}^{-3}\,{\rm yr}^{-1}$. Furthermore, since the sensitivities of aLIGO and ET are $h_{\rm rms}^{\rm LIGO}\sim 3\times 10^{-22}$ and $h_{\rm rms}^{\rm ET}\sim 2\times 10^{-23}$ for signals of $1-4\,$kHz respectively \citep[also see Figure 3 of][]{gao17}, from Eq. (16), we obtain the detection horizons of these detectors, $d_{\rm LIGO}\lesssim150\,$Mpc and $d_{\rm ET}\lesssim2.2\,$Gpc for $\alpha\sim 0.1$, and thus we find the detectable event rates, ${\cal R}_{\rm LIGO}=(4\pi/3)d_{\rm LIGO}^3{\cal R}_{\rm osc}\sim 20\,{\rm yr}^{-1}$ and ${\cal R}_{\rm ET}=(4\pi/3)d_{\rm ET}^3{\cal R}_{\rm osc}\sim 7\times 10^4\,{\rm yr}^{-1}$. For a small radial pulsation amplitude $\alpha\sim 0.03$, however, the detection horizons of the detectors decrease by a factor $\sim 3.3$, so the detectable event rates become ${\cal R}_{\rm LIGO}\sim 0.6\,{\rm yr}^{-1}$ and ${\cal R}_{\rm ET}\sim 2\times 10^3\,{\rm yr}^{-1}$. Therefore, it seems that next-generation detectors such as ET would be able to detect a large number of GW events from radially oscillating pulsar remnants every year.

Besides binary neutron star mergers, millisecond pulsars including millisecond magnetars are produced through the other astrophysical processes, for example, core collapse of massive stars, accretion-induced collapse of white dwarfs, and mergers of binary white dwarfs. These stars may have radial oscillations. Thus, our analysis in this paper can also be applied to such stars. Of course, the complete waveforms of GWs from binary neutron star mergers are different from those of GWs from the other processes, since the former include the GWs at the inspiraling stage and the latter have only GWs from the radial oscillations. This property could be used to distinguish between the binary neutron star merger and other processes.

We finally discuss the effects of $\gamma$. This parameter is actually dependent of the equation of state for neutron-star matter. For realistic equations of state, $\gamma$ is in the range of two to three \citep{haensel02}, so that $\eta_\gamma\sim 1-2.5$ and $\kappa_\gamma\sim 1-1.54$. Since $h_c$ is weakly dependent of $\eta_\gamma$ and independent of $\kappa_\gamma$, from Eq. (16), the characteristic strain of GWs is hardly affected by $\gamma$ directly. However, from Eqs. (3) and (7), the frequency (duration) of GWs increases (decreases) with increasing $\gamma$. Even so, as long as the stellar period is in the order of $\sim 1\,$ms and the surface field strength is not beyond $10^{16}\,$G, the GW damping timescale is much smaller than the viscous damping timescale and also the PMR damping timescale (even though these timescales are all dependent of the mass-radius relation). This conclusion is always true for any realistic equation of state of dense matter above the nuclear density, no matter whether the post-merger compact object is a neutron star or a strange quark star, and thus our analysis in Section 2 is valid.

\section{Summary}

We have proposed a model to explore GWs from post-merger radially oscillating, rapidly rotating pulsars after binary neutron star mergers, and found that rotation-induced gravitational radiation is the dominant damping mechanism of the radial oscillations. Some other conclusions are summarized below.

First, the resulting GWs have a frequency $f\simeq 4.1R_6^{-3/2}(M/2.5M_\odot)^{1/2}\,$kHz, which is very close to the cutoff frequency shown by numerical simulations, while the characteristic strain $h_c$ is not only independent of the neutron-star rotational period but also weakly dependent of the stellar radius. Furthermore, for the same source, our derived strain is much stronger than given by numerical simulations at a frequency $\gtrsim3$\,kHz for $\alpha\sim 0.03-0.1$. Therefore, the GWs from radially oscillating neutron star remnants would have a dominant contribution to the high-frequency spectrum.

Second, the estimated event rate is so high that the GW events would be detectable with the advanced LIGO/Virgo or next-generation detectors such as ET. If detected, such GW events, together with their frequencies, durations and strains, would be used to constrain the properties of neutron star remnants, for example, the mass-radius relation (from Eq. [3]), rotational periods (from Eq. [7]) and radial pulsation amplitudes (from Eq. [16]). If not detected, on the other hand, an upper limit on $\alpha$ would be given.

Third, for GW170817, no post-merger signal was found by the aLIGO/Virgo detectors. This requires that $\alpha\lesssim 0.6\sqrt{f_3}$, which seems too loose to provide any useful constraint on the neutron star remnant.

Finally, the other astrophysical formation processes of newborn radially oscillating, strongly magnetized millisecond pulsars include core collapse of massive stars, accretion-induced collapse of white dwarfs, and mergers of binary white dwarfs. Our model can also be applied to these newborn pulsars.

\section*{Acknowledgements}
The author would like to thank the referee for helpful comments and suggestions, and He Gao, Maurice H. P. M. van Putten, Yun-Wei Yu, and Bing Zhang for their discussions. This work is supported by the National Key Research and Development Program of China (grant No. 2017YFA0402600), the National Basic Research Program of China (``973 Program", grant No. 2014CB845800), and the National Natural Science Foundation of China (grant No. 11573014 and 11833003).

\label{lastpage}


\begin{thebibliography}{}
\bibitem[Abbott et al.(2017a)]{Abbott17a} Abbott, B.~P., Abbott, R., Abbott, T.~D., et al. 2017a, \prl, 119, 161101
\bibitem[Abbott et al.(2017b)]{Abbott17b} Abbott, B.~P., Abbott, R., Abbott, T.~D., et al. 2017b, \apjl, 848, L12
\bibitem[Abbott et al.(2017c)]{Abbott17c} Abbott, B.~P., Abbott, R., Abbott, T.~D., et al. 2017c, \apjl, 851, L16
\bibitem[Abbott et al.(2018)]{Abbott18} Abbott, B.~P., Abbott, R., Abbott, T.~D., et al. 2018, arXiv:1810.02581
\bibitem[Andersson(1998)]{andersson98} Andersson, N. 1998, \apj, 502, 708
\bibitem[Andersson \& Ho(2018)]{andersson18} Andersson, N., \& Ho, W. C. G. 2018, \prd, 97, 023016
\bibitem[Ai et al.(2018)]{ai18} Ai, S. K., Gao, H., Dai, Z. G., Wu, X. F., Li, A., Zhang, B., \& Li, M. Z. 2018, \apj, 860, 57
\bibitem[Arun et al.(2005)]{arun05} Arun, K. G., Iyer, B. R., Sathyaprakash, B. S., \& Sundararajan, P. A. 2005, \prd, 71, 084008
\bibitem[Baiotti \& Rezzolla(2017)]{baiotti17} Baiotti, L., \& Rezzolla, L. 2017, Rep. Prog. Phys., 80, 096901
\bibitem[Bauswein et al.(2018)]{bau18} Bauswein, A., Bastian, N. U. F., Blaschke, D. B., et al. 2018, arXiv:1809.01116
\bibitem[Bauswein \& Stergioulas(2015)]{bau15} Bauswein, A., \& Stergioulas, N. 2015, \prd, 91, 124056
\bibitem[Berger(2014)]{berger14} Berger, E. 2014, ARA\&A, 52, 43
\bibitem[Bernuzzi et al.(2015)]{ber15} Bernuzzi, S., Dietrich, T., \& Nagar, A. 2015, \prl, 115, 051101
\bibitem[Bondarescu et al.(2009)]{Bond09} Bondarescu, R., Teukolsky, S. A., \& Wasserman, I. 2009, \prd, 79, 104003
\bibitem[Bondarescu \& Wasserman(2013)]{Bond13} Bondarescu, \& Wasserman, I. 2013, \apj, 778, 9
\bibitem[Cameron(1965)]{cameron65} Cameron, A. G. W. 1965, \nat, 205, 787
\bibitem[Chau(1967)]{chau67} Chau, W. Y. 1967, \apj, 147, 664
\bibitem[Chatziioannou et al.(2017)]{chat17} Chatziioannou, K., Clark, J. A., Bauswein, A., et al. 2017, \prd, 94, 124035
\bibitem[Cheng \& Dai(1998)]{cheng98} Cheng, K. S., \& Dai, Z. G. 1998, \apj, 492, 281
\bibitem[Corsi \& M{\'e}sz{\'a}ros(2009)]{Corsi09} Corsi, A., \& M{\'e}sz{\'a}ros, P. 2009, \apj, 702, 1171
\bibitem[Cox(1980)]{cox80} Cox, J. P. 1980, Theory of Stellar Pulsation (Princeton University Press), Chapter 8
\bibitem[Dai(2004)]{Dai04} Dai, Z.~G. 2004, \apj, 606, 1000
\bibitem[Dai \& Lu(1996)]{dai96} Dai, Z.~G., \& Lu, T. 1996, Z. Phys. A, 355, 415
\bibitem[Dai \& Lu(1998a)]{Dai98a} Dai, Z.~G., \& Lu, T. 1998a, \aap, 333, L87
\bibitem[Dai \& Lu(1998b)]{Dai98b} Dai, Z. G., \& Lu, T. 1998b, \prl, 81, 4301
\bibitem[Dai et al.(2006)]{Dai06} Dai, Z.~G., Wang, X.~Y., Wu, X.~F., \& Zhang, B. 2006, Science, 311, 1127
\bibitem[Dai et al.(2016)]{Dai16} Dai, Z.~G., Wang, S.~Q., Wang, J.~S., Wang, L.~J., \& Yu, Y.~W. 2016, \apj, 817, 132
\bibitem[Dall'Osso et al.(2011)]{Dall11} Dall'Osso, S., Stratta, G., Guetta, D., Covino, S., De Cesare, G., \& Stella, L. 2011, \aap, 526, A121
\bibitem[Dall'Osso et al.(2015)]{Dall15} Dall'Osso, S., Giacomazzo, B., Perna, R., \& Stella, L. 2015, \apj, 789, 25
\bibitem[Duncan(1989)]{duncan89} Duncan, R. C. 1989, \nat, 340, 699
\bibitem[Gao et al.(2017)]{gao17} Gao, H., Cao, Z., \& Zhang, B. 2017, \apj, 844, 112
\bibitem[Gao et al.(2016)]{gao16} Gao, H., Zhang, B., \& L{\"u}, H. J. 2016, \prd, 93, 044065
\bibitem[Geng et al.(2018)]{geng18} Geng, J. J., Dai, Z. G., Huang, Y. F., Wu, X. F., Li, L. B., Li, B., \& Meng, Y. Z. 2018, \apjl, 856, L33
\bibitem[Giacomazzo \& Perna(2013)]{Giacomazzo13} Giacomazzo, B., \& Perna, R. 2013, \apjl, 771, L26
\bibitem[Heintzmann \& Nitsch(1972)]{heintz72} Heintzmann, H., \& Nitsch, J. 1972, \aap, 21, 291
\bibitem[Haensel et al.(2002)]{haensel02} Haensel, P., Levenfish, K. P., \& Yakovlev, D. G. 2002, \aap, 394, 213
\bibitem[Haensel et al.(2009)]{haensel09} Haensel, P., Zdunik, J. L., Bejger, M., \& Lattimer, J. M. 2009, \aap, 502, 605
\bibitem[Ho \& Lai(2000)]{ho00} Ho, W. C. G., \& Lai, D. 2000, \apj, 543, 386
\bibitem[Hoyle et al.(1964)]{hoyle64} Hoyle, F., Nalikar, J. V., \& Wheeler, J. A. 1964, \nat, 203, 914
\bibitem[Li et al.(2016)]{li16} Li, A., Zhang, B., Zhang, N. B., Gao, H., Qi, B., \& Liu, T. 2016, \prd, 94, 083010
\bibitem[Li et al.(2018)]{li18} Li, S. Z., Liu, L. D., Yu, Y. W., \& Zhang, B. 2018, \apjl, 861, L12
\bibitem[Madsen(1992)]{madsen92} Madsen, J. 1992, \prd, 46, 3290
\bibitem[Madsen(2000)]{madsen00} Madsen, J. 2000, \prl, 85, 10
\bibitem[Metzger et al.(2018)]{metzger18} Metzger, B. D., Thompson, T. A., \& Quataert, E. 2018, \apj, 856, 101
\bibitem[Mishra et al.(2010)]{mishra10} Mishra, C. K., Arun, K. G., Iyer, B. R., \& Sathyaprakash, B. S. 2010, \prd, 82, 064010
\bibitem[Most et al.(2018)]{most18} Most, E. R, Papenfort, L. J., Dexheimer, V., et al. 2018, arXiv:1807.03684
\bibitem[Nakar(2007)]{nakar07} Nakar, E. 2007, Phys. Rep., 442, 166
\bibitem[Owen et al.(1998)]{owen98} Owen, B. J., Lindblom, L., Cutler, C., et al. 1998, \prd, 58, 084020
\bibitem[Passamonti et al.(2013)]{pas13} Passamonti, A., Gaertig, E., Kokkotas, K. D., \& Doneva, D. 2013, \prd, 87, 084010
\bibitem[Piro et al.(2017)]{Piro17} Piro, A. L., Giacomazzo, B., \& Perna, R. 2017, \apjl, 844, L19
\bibitem[Piro et al.(2019)]{piro18} Piro, L., Troja, E., Zhang, B., et al. 2019, \mnras, 483, 1912
\bibitem[Reisenegger \& Goldreich(1992)]{Reisen92} Reisenegger, A., \& Goldreich, P. 1992, \apj, 395, 240
\bibitem[Rowlinson et al.(2010)]{Rowlinson10} Rowlinson, A., O'Brien, P. T., Tanvir, N. R., et al. 2010, \mnras, 409, 531
\bibitem[Rowlinson et al.(2013)]{Rowlinson13} Rowlinson, A., O'Brien, P.~T., Metzger, B.~D., Tanvir, N.~R., \& Levan, A.~J. 2013, \mnras, 430, 1061
\bibitem[Sawyer(1980)]{sawyer80} Sawyer, R. F. 1980, \apj, 237, 187
\bibitem[Sawyer(1989)]{sawyer89} Sawyer, R. F. 1989, \prd, 39, 3804
\bibitem[Sawyer \& Soni(1979)]{sawyer79} Sawyer, R. F., \& Soni, A. 1979, \apj, 230, 859
\bibitem[Shapiro \& Teukolsky(1983)]{Shapiro83} Shapiro, S.~L., \& Teukolsky, S.~A. 1983, Black Holes, White Dwarfs, and Neutron Stars: The Physics of Compact Objects (New York: Wiley), Chapter 16
\bibitem[Shibata \& Kiuchi(2017)]{shibata17} Shibata, M., \& Kiuchi, K. 2017, \prd, 95, 123003
\bibitem[Sun et al.(2015)]{sun15} Sun, B., Cao, Z., Wang, Y., \& Yeh, H.-C. 2015, \prd, 92, 044034
\bibitem[Takami et al.(2015)]{tak15} Takami, K., Rezzolla, L., \& Baiotti, L. 2016, \prd, 91, 064001
\bibitem[Thorne(1987)]{thorne87} Thorne, K. S. 1987, in Three Hundred Years of Gravitation, edited by S. W. Hawking \& W. Israel (Cambridge: Cambridge Univ. Press), 330
\bibitem[Timmes et al.(1996)]{timmes96} Timmes, F. X., Woosley, S. E., \& Weaver, T. A. 1996, \apj, 457, 83
\bibitem[Torres-Rivas et al.(2018)]{torres18} Torres-Rivas, A., Chatziioannou, K., Bauswein, A., \& Clark, J. A. 2018, \prd, submitted, arXiv:1811.08931
\bibitem[van Putten \& Della Valle(2019)]{vanPutten18} van Putten, M. H. P. M., \& Della Valle, M. 2019, \mnras, 482, L46
\bibitem[Villar et al.(2017)]{villar17} Villar, V. A., Guillochon, J., Berger, E., et al. 2017, \apjl, 851, L21
\bibitem[Weinberg et al.(2013)]{Wein13} Weinberg, N. N., Arras, P., \& Burkart, J. 2013, \apj, 769, 121
\bibitem[Yu et al.(2018)]{yu18} Yu, Y. W., Liu, L. D., \& Dai, Z. G. 2018, \apj, 861, 114
\bibitem[Zappa et al.(2018)]{zappa18} Zappa, F., et al. 2018, \prl, 120, 111101
\bibitem[Zhang \& M{\'e}sz{\'a}ros(2001)]{zhang01} Zhang, B., \& M{\'e}sz{\'a}ros, P. 2001, \apjl, 552, L35
\end{thebibliography}
\end{document}